\documentclass[11pt,a4wide]{article}

\usepackage{amsfonts}
\usepackage{amsbsy}
\usepackage{epsfig}
\usepackage{latexsym}
\input amssym.def
\input amssym.tex

\advance \topmargin by -\headheight
\advance \topmargin by -\headsep
\evensidemargin \oddsidemargin
\advance\hoffset by -3mm  
\advance\voffset by  8mm  
\def\bbbc{{\mathchoice {\setbox0=\hbox{$\displaystyle\rm C$}\hbox{\hbox
to0pt{\kern0.4\wd0\vrule height0.9\ht0\hss}\box0}}
{\setbox0=\hbox{$\textstyle\rm C$}\hbox{\hbox
to0pt{\kern0.4\wd0\vrule height0.9\ht0\hss}\box0}}
{\setbox0=\hbox{$\scriptstyle\rm C$}\hbox{\hbox
to0pt{\kern0.4\wd0\vrule height0.9\ht0\hss}\box0}}
{\setbox0=\hbox{$\scriptscriptstyle\rm C$}\hbox{\hbox
to0pt{\kern0.4\wd0\vrule height0.9\ht0\hss}\box0}}}}

\makeatletter
\@addtoreset{equation}{section}
\makeatother

\begin{document}

\hfuzz=100pt \title{{\Large \bf{Embedding FLRW Geometries in Pseudo-Euclidean and Anti-de Sitter Spaces}}}
\author{\\M M Akbar\footnote{E-mail: akbar@utdallas.edu} \\
Department of Mathematical Sciences,\\ University of Texas at Dallas, \\ Richardson, TX 75080, USA} \date{\today} \maketitle
\begin{abstract}
  \noindent
Contrary to the general consensus in the literature that Friedmann--Lema\^{i}tre--Robertson--Walker (FLRW) geometries are of embedding class one (i.e.,\ embeddable in one higher dimensional pseudo-Euclidean spaces), we show that the most general $k=0$ and $k=-1$ FLRW geometries are of embedding class two, and their corresponding pseudo-Euclidean spaces have strictly one and two negative eigenvalues, respectively. These are particular results that follow from the new perspective on FLRW embedding that we develop in this paper, namely that these embeddings are equivalent to unit-speed parametrized curves in two or three dimensions. A careful analysis of appropriate tensor fields then gives identical results and further explains why the class-two geometries remained hidden. However, the signatures of the embedding spaces, as well as the explicit embedding formulae, follow only from the curve picture. This also streamlines the comparatively difficult $k=0$ class and provides new explicit embedding formulae for it and reproduces known embedding formulae for the $k=1,-1$ classes. Embedding into anti-de Sitter space in one higher dimension can likewise be done by constructing associated curves. In particular, we find that all $k=1$ and mildly restricted subclasses of $k=0, -1$ geometries are embeddable in anti-de Sitter space in one higher dimension.
\end{abstract}
\section{Introduction}
It is common practice in general relativity to introduce students to the de Sitter and anti-de Sitter (AdS) spacetimes as ``hyperboloids" embedded in one-dimensional higher flat spaces with negative signatures, much like the sphere in Euclidean space. Unlike the Schwarzschild spacetime, which is usually presented as the first nontrivial solution of the Einstein equations, these two spaces are of constant curvature and their embeddings help build intuition and develop an appreciation for the benefit of using different coordinates (see, for example, \cite{HE}). At this point, students are not introduced to various embedding theorems of differential geometry and one pays little attention to the different signatures of the embedding flat spaces.

Riemannian geometry studies geometric objects intrinsically. Nonetheless, embeddings of Riemannian spaces, local or global, into higher dimensional flat or nonflat spaces can provide additional insights, and their study constitutes a well-researched area of mathematics with many classic results \cite{Besse}. Many embedding results in the Riemannian case have pseudo-Riemannian counterparts \cite{EisenhartRiem, Friedman1965}. These include the extended Schl\"{a}fli--Janet--Carter--Burstin theorem that any $d$-dimensional pseudo-Riemannian space can be locally isometrically embedded in a $d(d+1)/2$-dimensional pseudo-Euclidean space.\footnote{See \cite{Pavsic:2000qy} for a historical account and references.} For a particular $d$-dimensional geometry, especially in the presence of symmetry, the dimension of the pseudo-Euclidean space could be as low as $(d+1)$. Of course, a necessary condition for any embedding is that the pseudo-Euclidean space (or the embedding geometry) cannot have fewer negative or positive eigenvalues than the embedded geometry. The number of extra dimensions in the ``minimal embedding" of a metric is what constitutes its ``embedding class." Thus, every four-dimensional Lorentzian geometry has a (fixed) embedding class ranging between one and six ($p=1$ to $p=6$). The constant curvature spaces above, for example, are of embedding class one and the Schwarzschild solution is of class two. Embedding classes provide an invariant classification scheme, which can be used in conjunction with the other classification schemes based on symmetry and the Petrov type for solutions in general relativity \cite{ExactSol2003}.

In this paper, we will consider local isometric embedding of Friedmann--Lema\^{i}tre--Robertson--Walker (FLRW) spacetimes in pseudo-Euclidean spaces, and then in AdS space. Explicit embedding formulae for these spacetimes in the Minkowski space were first obtained by Robertson \cite{Robertson:1928, Robertson:1933zz}. These were derived independently by Rosen, who constructed pseudo-Euclidean embeddings of many other known spacetimes and was unaware of Robertson's work \cite{Rosen1965}. Interestingly, the same formulae kept being rediscovered in various contexts until fairly recently, sometimes with erroneous or premature conclusions; see \cite{GulamovSmolyakov2011} for a comprehensive list of references.\footnote{The only additional references we were able to locate are \cite{Nielsen} and the later work \cite{Paston:2012pb}, which also derive identical embedding formulae.} In any case, the collective summary is as follows: all possible $k=0,1$ FLRW geometries, but only a certain type of the $k=-1$ (e.g. those obeying the Einstein equations of the standard cosmological model) are embeddable in one-dimensional higher Minkowski space; the other $k=-1$ subclass of geometries is embeddable in one-dimensional higher pseudo-Euclidean space of two negative eigenvalues.

Our initial purpose of revisiting FLRW embedding further was to offer a simple geometric picture in which these embeddings can be seen as parametrized unit-speed curves in an auxiliary two-dimensional Minkowski space. Known embedding formulae for $k=1$ and $k=-1$ follow from their respective curves in this picture, and the comparatively tenacious $k=0$ class is brought into an equal footing with the $k=1, -1$ classes and acquires embedding formulae that are more intuitive. However, looking carefully at the curve construction, we noticed that two possible types of geometries (belonging to the $k=0$ and $k=-1$ classes) went unnoticed in all previous studies and that their embedding curves can be constructed only by allowing one additional dimension (and appropriate signatures), making them embedding class two. This makes us consider the more traditional tensor construction for embedding, which agrees in every detail with the curve construction and, in addition, provides a plausible explanation as to why such geometries remained unnoticed. With these two classes taken into account, embeddings for all possible Lorentzian FLRW geometries are covered.

We then exploit the picture of the curve in determining embeddings of FLRW spacetimes in the AdS space. The particular advantage of classification based on pseudo-Euclidean embeddings disappears in nonflat embeddings. However, nonflat embeddings are mathematically and physically interesting, tracing their roots back to the original work of Kaluza--Klein, who showed that five-dimensional vacuum general relativity gives rise to four-dimensional general relativity with matter sources. More recently, finding realistic four-dimensional cosmological ``brane" universe models embedded in higher dimensional ``bulk" spacetimes has been a constant theme (see, for example, \cite{Andrianopoli:1999kx}, and references therein). In particular, in various supergravity, string and M theories, the AdS space serves as the bulk. Despite this interest, a geometric embedding of FLRW in the AdS space, without a contribution from an energy--momentum tensor or otherwise, has not been done before.
\section{FLRW in Pseudo-Euclidean Spaces}
A $d$-dimensional FLRW geometry is a warped product of a $(d-1)$-dimensional fiber of constant scalar curvature $k$ with a one-dimensional base:
\begin{eqnarray}
ds^2&=&\pm d\tau^2+ a(\tau)^2d\Omega^2_{d-1,k} , \label{FLRWCos}\\
d\Omega^2_{d-1,k}&=& \left \{ \begin{array}{ll} g(\mathbb{S}^{d-1},{\rm
can})\ ,& k=1\ ,\\ g(\mathbb{E}^{d-1},{\rm can})\ ,& k=0\ ,\\
g(\mathbb{H}^{d-1},{\rm can})\ ,& k=-1\ . \end{array} \right.\label{FLRWCos1}
\end{eqnarray}
Because of the maximal symmetry of the fiber, it is sufficient to work with $d=4$ without loss of generality. We will consider only the Lorentzian case below, from which one should be able to work out the Riemannian case with little modification, mostly by changing signs. We will refer to the warping function $a(\tau)$ as ``scale factor" following the physics literature, and will assume it to be continuous and differentiable.

It is customary to write the four-dimensional (\ref{FLRWCos})--(\ref{FLRWCos1}) in the following combined form:
\begin{equation}
ds^2=-d\tau^2+a(\tau)^2\left(\frac{dr^2}{1-kr^2} +r^2 (d\theta^2+ \sin^2\theta d\phi^2)\right),\label{CosmoCustomary}
\end{equation}
from which the above, and other forms can be worked out with simple transformations of coordinates. With the ansatz (\ref{FLRWCos})--(\ref{FLRWCos1}), or (\ref{CosmoCustomary}), the Einstein equations reduce to a set of ordinary differential equations in $a(\tau)$, the solution of which depends on the matter content and the scalar curvature $k$ of the (homogeneous) hypersurface. However, we will consider all possible geometries of the form (\ref{FLRWCos})--(\ref{FLRWCos1}) that do not necessarily follow from any Einstein equations; however, the standard cosmological cases will be considered alongside as a special class.\footnote{Following the growing practice in the field, we use FLRW instead of FRW. Also, strictly, when one is not using the Einstein equations, one should refer to the geometry as RW. In our case, any reference to the Einstein equations will be made explicitly and we will use FLRW generally, as defined above.}

As a precursor to the particular approach of embedding that we will develop, we revisit the embedding of de Sitter space in one-dimensional higher Minkowski space to see how this can be studied as a curve in a two-dimensional Minkowski space. We start with the Minkowski metric in polar coordinates in five dimensions:
\begin{equation}
ds^2= -dt^2+dr^2+r^2 d\Omega_{3,1}^2 \label{MinkowskiPolar}
\end{equation}
where $d\Omega_{3,1}^2$ is the standard $SO(4)$-invariant metric on $\mathbb{S}^3$ as above. As usual with polar coordinates, $r=0$ is a coordinate singularity of this metric. Now, consider an $SO(4)$-invariant hypersurface in ({\ref{MinkowskiPolar}}) given by
\begin{equation}
t=F(r).\label{hyper}
\end{equation}
This represents a curve in the $t$--$r$  plane of ({\ref{MinkowskiPolar}}), to each point of which a sphere of radius equal to the curve's $r$ coordinate is attached. The induced metric on this four-dimensional hypersurface is then
\begin{equation}
ds^2=(1-F'^2)dr^2+r^2 d\Omega_{3,1}^2.
\end{equation}
This is timelike if $F'^2 > 1$. Taking $(F'^2-1)= 1/ (r^2-1)$, one obtains
\begin{equation}
ds^2=-\frac{1}{r^2-1}dr^2+r^2 d\Omega_{3,1}^2,\label{deSitter0}
\end{equation}
which is easily recognizable as the de Sitter metric. With $r=\cosh{\tau}$, it can be turned into the more familiar form:
\begin{equation}
ds^2=-d\tau^2+ \cosh^2{\tau} d\Omega_{3,1}^2,\label{deSitter1}
\end{equation}
which covers the whole of de Sitter space except at the same trivial singularities of (\ref{MinkowskiPolar}). Finally, integrating for $F(r)$, one obtains
\begin{equation}
t=\pm \sqrt{r^2-1}
\end{equation}
as the hypersurfaces (\ref{hyper}). These are nothing but the two sides of the hyperbola $r^2-t^2=1$ in the $t$--$r$  subspace of ({\ref{MinkowskiPolar}}). The transition from the usual picture of de Sitter as a hyperboloid in one-dimensional higher Minkowski space to that of a curve in two-dimensional Minkowski space is possible because we were able to identify, and factor out, the common spherical symmetry between the Minkowski and de Sitter spaces by writing them in similar coordinates, (\ref{MinkowskiPolar}) and (\ref{deSitter0}), respectively. In both, $r$ plays the common role of the radius of the sphere. This particular advantage disappears as we consider FLRW geometries, but can be circumvented by moving into a parametrized description of curves, as we will see below.
\subsection{The $k=1$ FLRW}
We will write the $k=1$ FLRW spacetime metric in the following warped form:
\begin{equation}
ds^2= -d\tau^2+ a(\tau)^2 \left(d\chi^2+ \sin^2{\chi}(d\theta^2+\sin^2{\theta} d\phi^2)\right)\label{FLRWk1}
\end{equation}
and take the five-dimensional Minkowski metric as before:
\begin{equation}
ds^2= -dt^2+dr^2+r^2 \left(d\chi^2+ \sin^2{\chi}(d\theta^2+\sin^2{\theta} d\phi^2)\right). \label{5dMink}
\end{equation}
We can proceed as in the example above by taking an $SO(4)$-invariant hypersurface defined by $t=F(r)$, which represents a curve in the $t$--$r$ subspace. However, this would not, in general, lead to a compact expression as was possible for the de Sitter space. A better approach would be to consider the curve as parametrized with $\eta$ (see, for example, \cite{DoCarmo, ONeill1, ONeill}):
\begin{equation}
{\cal{C}}=(b(\eta), a(\eta)).\label{konecurve}
\end{equation}
Note that this means we have taken $r=a(\eta)$ and $t=b(\eta)$. The $4-d$ metric on the hypersurface is then
\begin{equation}
ds^2= (-b'^2 + a'^2) d\eta^2+ a(\eta)^2 \left(d\chi^2+ \sin^2{\chi}(d\theta^2+\sin^2{\theta} d\phi^2)\right), \label{5dMink1}
\end{equation}
where prime represents differentiation with respect to $\eta$. If we choose
\begin{equation}
-b'^2 + a'^2=-1 \label{cond1}
\end{equation}
the metric (\ref{5dMink1}) gives (\ref{FLRWk1}) upon identifying $\tau$ and $\eta$. Thus, the curve (\ref{konecurve}) in the $t$--$r$ plane -- which is a two-dimensional Minkowski space -- is a unit-speed timelike curve. Rearranging, we find
\begin{equation}
b'(\tau)^2= {1 + a'(\tau)^2}, \label{cond11}
\end{equation}
which for any $a(\tau)$ can be integrated. Thus, for any $a(\tau)$, there exits the unit-speed curve parametrized by $\tau$:
\begin{equation}
{\cal{C}}: \tau \rightarrow \left( \int \sqrt{1+ a'(\tau)^2}\,\, d\tau, a(\tau)\right). \label{konecurveexplicit}
\end{equation}
Since the coordinates of the spheres are already identified, to see if we indeed have an isometric embedding or not, we need to check the curve as a map from the real line. It immediately follows from (\ref{cond1}) that this is, first of all, an immersion since $b'$ and $a'$ cannot be simultaneously zero, and so the Jacobian of the map will have the necessary rank. In addition, the curve (\ref{konecurveexplicit}) is one-to-one for any continuous and differentiable $a(\tau)$; thus, it is an embedding.

The above result for $k=1$ is geometric and does not presuppose any field equations (cf.\ the $k=-1$ class below), so it holds in general. This applies to cosmological and wormhole-type solutions alike. Note that all cosmological solutions in general relativity under reasonable matter assumptions will have an initial singularity. Thus, $\tau=0$ will be a true (general relativistic) singularity, which will coincide with the $r=0$ (removable) coordinate singularity of embedding metric (\ref{5dMink}).

The explicit embedding formula for $k=1$ (see, for example, \cite{Rosen1965}) has as the extra-dimensional coordinate:
\begin{equation}
z_5=\int \sqrt{1+ a'(\tau)^2}\, d\tau,\label{exradimcoor1}
\end{equation}
which easily follows from (\ref{cond11}) above. Note that (\ref{cond11}) was explicitly derived in \cite{GulamovSmolyakov2011}. However, its interpretation as a unit-speed parametrized curve was not noted. This interpretation will be most fruitful in determining the embedding of the most general $k=-1$ and $k=0$ geometries and reformulating the $k=0$ class.
\subsubsection{de Sitter as a $k=1$ model}
Looking at it as a FLRW spacetime, the de Sitter metric (\ref{deSitter1}) is a $k=1$ model with $a(\tau)= \cosh{\tau}$.
The parametrized curve (\ref{konecurveexplicit}) is then
\begin{equation}
{\cal{C}}=\left(t(\tau), r(\tau) \right)=\left(\cosh{\tau}, \sinh{\tau} \right),\label{desitk1embed}
\end{equation}
which gives the same hyperbola $r^2-t^2=1$ upon the elimination of $\tau$ found above. Note that de Sitter can be written in coordinates that turn it into a $k=0$ model. We will discuss this in Section \ref{sec:Desitterzero}. On a historical note, de Sitter was the first cosmological model Robertson considered for embedding \cite{Robertson:1928}.
\subsection{The $k=-1$ FLRW}
Analogous to (\ref{FLRWk1}), we will work with the FLRW $k=-1$ metric in the following form:
\begin{equation}
ds^2= -d\tau^2+ a(\tau)^2 \left(d\chi^2+ \sinh^2{\chi}(d\theta^2+\sin^2{\theta} d\phi^2)\right),\label{FLRWk2}
\end{equation}
and with the following, somewhat less popular, form of the Minkowski metric:
\begin{equation}
ds^2= -dt^2+ dr^2+ t^2 \left(d\chi^2+ \sinh^2{\chi}(d\theta^2+\sin^2{\theta}\phi^2)\right).\label{MinkLoCone5d}
\end{equation}
As before, the curve in the $t$--$r$ plane\footnote{From now on we do not introduce the auxiliary parameter $\eta$ since we will finally identify it with $\tau$.}
\begin{equation}
{\cal{C}}=(a(\tau), b(\tau))
\end{equation}
would represent the following hypersurface:
\begin{equation}
ds^2= (-a'^2 + b'^2) d\tau^2+ a(\tau)^2 \left(d\chi^2+ \sinh^2{\chi}(d\theta^2+\sin^2{\theta}\phi^2)\right),
\end{equation}
which gives the metric (\ref{FLRWk2}) provided
\begin{equation}
b'^2=a'^2-1. \label{condminusk}
\end{equation}
It is easy to see, using similar arguments as the $k=1$ case, that this is an isometric embedding. As before, (\ref{condminusk}) gives the extra-dimensional coordinate for the Minkowski that appears in the literature:
\begin{equation}
z_5=\int \sqrt{a'(\tau)^2-1}\, d\tau.\label{exradimcoor2}
\end{equation}
Note that (\ref{condminusk}), or (\ref{exradimcoor2}), clearly requires $a'^2 > 1$ for the scale factor. The embedding of $a'^2 < 1$ models can be achieved by letting $z_5\rightarrow iz_5$, i.e., by flipping the sign of the $r$ coordinate in (\ref{MinkLoCone5d}). Thus, this would require a pseudo-Euclidean space of two timelike directions -- just as in the well-known AdS space, a $k=-1$ model with $a(\tau)=\cos(\tau)$. The Einstein field equations for the standard $k=-1$ FLRW cosmological models driven by a perfect fluid (see, for example, \cite{WaldBook}) imply
\begin{equation}
a'^2=\frac{8 \pi}{3} \rho a^2 +3,\label{EFF}
\end{equation}
which guarantees $a'^2 > 1$. Thus, within the standard general relativistic cosmological context, one can rule out $a'^2 < 1$. We note that the only work in the literature with an argument analogous to this is \cite{GulamovSmolyakov2011}, which also concurrently showed the requirement for two timelike directions for the $a'^2 < 1$ subclass. It seems that others did not pay much attention to the condition coming from (\ref{exradimcoor2}), since either they did not mention the field equations to justify $a'^2 > 1$ or they did not discuss the need for two timelike directions for the $a'^2 < 1$ subclass despite the well-known example of the anti-de Sitter spacetime (see, for example, \cite{LachiezeRey:2000my}).\footnote{Rosen's embedding formulae in \cite{Rosen1965} had the field equations alongside, but there was no comment on the embedding of the $a'^2 < 1$ subclass.} In any case, to summarize, for $k=-1$ geometries, there is a sharp dividing line between the $a'^2 > 1$ and $a'^2 < 1$ subclasses in terms of the pseudo-Euclidean spaces they can be embedded in.
\subsubsection{Minimal embedding of the general $k=-1$ FLRW geometries}
In all works with $k=-1$, what seems to have been missed altogether are those geometries that cross between the two subclasses, $a'>1$ and $a'<1$. Take, for example, $a(\tau)=e^{\tau/2}$. This is nonsingular and there is no mathematical reason to exclude geometries like this. What would be the minimal pseudo-Euclidean space for them? As we will see below, the construction of a unit-speed curve leads to a definitive answer to this question.
\\
\\
{\bf{Theorem 2.1:}} The general $k=-1$ FLRW geometry is of embedding class two ($p=2$) with strictly two negative eigenvalues.
\\
\\
{\bf{Proof:}} It suffices to prove it in four dimensions, as before. Recall that the embedding pseudo-Euclidean space must have at least the same number of positive or negative eigenvalues. This gives $\mathbb{E}^{3,2}$ or $\mathbb{E}^{4,1}$ as the only possible spaces to be considered in five dimensions. As we have already noted, $a'^2 \ge 1$ and $a'^2 \le 1$ cannot both be embedded in $\mathbb{E}^{3,2}$ or $\mathbb{E}^{4,1}$. This forces us to consider the six-dimensional pseudo-Euclidean space.
The eigenvalues then allow only three possibilities: $\mathbb{E}^{5,1}$, $\mathbb{E}^{4,2}$, and $\mathbb{E}^{3,3}$.
Starting with the metric on $\mathbb{E}^{4,2}$,
\begin{equation}
ds^2= -dt^2+ dr^2 -d\rho^2+ t^2 \left(d\chi^2+ \sinh^2{\chi}(d\theta^2+\sin^2{\theta}\phi^2)\right),\label{e42}
\end{equation}
the three-dimensional curve
\begin{equation}
{\cal{C}}=(a(\tau), b(\tau), c(\tau))
\end{equation}
gives the hypersurface,
\begin{equation}
ds^2= (-a'^2 + b'^2-c'^2) d\tau^2+ a(\tau)^2 \left(d\chi^2+ \sinh^2{\chi}(d\theta^2+\sin^2{\theta}\phi^2)\right),
\end{equation}
which is a Lorentzian $k=-1$ FLRW geometry with scale factor $a(\tau)$ provided
\begin{equation}
b'^2-c'^2= a'^2 -1.\label{generakneg}
\end{equation}
It is easy to see that this can always be satisfied for any given $a(\tau)$ in multiple ways due to the arbitrariness of $b(\tau)$ and $c(\tau)$. In particular, if one chooses $c(\tau)=\tau$, one gets $b(\tau)=a(\tau)$, for any $a(\tau)$.

Flipping (separately) the signs of the $\rho$ and $r$ coordinates in (\ref{e42}), the analogs of (\ref{generakneg}) in $\mathbb{E}^{5,1}$ and $\mathbb{E}^{3,3}$ are
\begin{equation}
b'^2+c'^2= a'^2 -1\label{generakneg1}
\end{equation}
and
\begin{equation}
b'^2-c'^2= 1-a'^2,\label{generakneg2}
\end{equation}
respectively. Neither of these allows an unrestricted $a(\tau)$ for any real choice of $b(\eta)$ and $c(\eta)$. Thus, $\mathbb{E}^{4,2}$ is the minimal embedding space for the most general $k=-1$ FLRW geometries. $\Box$

Note that this also shows that the embedding is nonrigid in that it is not unique due to the freedom of choice for $c(\eta)$ or $b(\eta)$.
\subsection{The General $k=0$ FLRW}
For the $k=0$ FLRW models, the spacetime ${\cal{M}}^4$ is $\mathbb{R}\times \mathbb{E}^3$ with metric:
\begin{equation}
ds^2=-dt^2 + a(\tau)^2(dx^2+dy^2+dz^2).\label{FLRWk3}
\end{equation}
To proceed in parallel with $k=\pm 1$, we need to write the Minkowski metric as a cone over three-dimensional  flat Euclidean space. Although not well known these days, such a form has existed for quite some time (see, for example, \cite{Robinson:1987mr}) and reads:
\begin{equation}
ds^2=-dt^2+dr^2+(t+r)^2(dx^2+dy^2+dz^2).\label{MinkNullCone5d}
\end{equation}
This is a null cone over $\mathbb{E}^{3}$ and one can check that its Riemann curvature tensor indeed is zero. As before, this does not cover the whole of Minkowski, but would suffice for embedding (\ref{FLRWk3}). The rest of the procedure is similar with little modification. We take the curve
\begin{equation}
{\cal{C}}=(b(\tau), c(\tau))
\end{equation}
in the $t$--$r$ plane. The corresponding hypersurface in the five-dimensional Minkowski is then
\begin{equation}
ds^2= (-b'^2 + c'^2) d\eta^2+ (b + c)^2 (dx^2+dy^2+dz^2).
\end{equation}
Taking
\begin{eqnarray}
b + c&=&a \label{k0curve1}\\
b'^2 - c'^2&=&1, \label{k0curve2}
\end{eqnarray}
we have the embedding. Differentiating (\ref{k0curve1}) and solving simultaneously with (\ref{k0curve2}), one obtains
\begin{eqnarray}
b'&=&\,\frac{1}{2}{\frac {{a'}^{2}-1}{a'}}\\
c'&=&\frac{1}{2}\,{\frac {{a'}^{2}+1}{a'}},
\end{eqnarray}
giving
\begin{equation}
{\cal{C}}=\left(\int \frac{1}{2}{\frac {{a'}^{2}-1}{a'}} d\tau, \int \frac{1}{2}\,{\frac {{a'}^{2}+1}{a'}} d\tau \right)\label{k0curve}
\end{equation}
and the following embedding in the Minkowski metric (\ref{MinkNullCone5d}):
\begin{eqnarray}
t&=& \int \frac{1}{2}{\frac {{a'}^{2}-1}{a'}} d\tau,\\
r&=& \int \frac{1}{2}\,{\frac {{a'}^{2}+1}{a'}} d\tau,\\
x &=& x,\\
y &=& y,\\
z &=& z.
\end{eqnarray}
This is clearly simpler than the set found by others (see, for example, \cite{Rosen1965}):
\begin{eqnarray}
z_1&=& \frac{1}{2}a(\chi^2+1)+\int \frac{dt}{2a'},\\
z_2&=& \frac{1}{2} a(\chi^2-1)+\int \frac{dt}{2a'},\\
z_3&=& a\,\chi \cos{\theta}, \\
z_4&=& a\,\chi \sin{\theta} \cos{\phi},\\
z_5&=& a\,\chi \sin{\theta} \sin{\phi},
\end{eqnarray}
which uses the Cartesian coordinates, $ds^2=-(dz_1)^2+(dz_2)^2+(dz_3)^2+(dz_4)^2+(dz_5)^2$. Note that, in this too, one is essentially expressing parts of the metric in polar coordinates.
\subsubsection{de Sitter as a $k=0$ model}\label{sec:Desitterzero}
As an illustration of the above, we consider the de Sitter metric in the following coordinates (see, for example, \cite{HE}):
\begin{equation}
ds^2=-d\tau^2 + e^{2\tau}(dx^2+dy^2+dz^2), \label{desitk0embed}
\end{equation}
for which $a(\tau)=e^{\tau}$. It is easy to check that the corresponding curve (\ref{k0curve}) in the $t$--$r$ plane is
\begin{equation}
{\cal{C}}=\left(\cosh{\tau}, \sinh{\tau} \right),
\end{equation}
which is in precise agreement with (\ref{desitk1embed}).
\subsubsection{Minimal embedding of the general $k=0$ FLRW geometries}
For $a'=0$, (\ref{k0curve1}) implies $b'=-c'$, which contradicts (\ref{k0curve2}). Thus, the analysis above is valid as long as $a'\ne 0$, i.e., in the absence of any local extremum of the scale factor. The $a(\tau)=\mathrm{const.}$ case is simple -- (\ref{FLRWk3}) is then just the Minkowski metric and can be embedded in the five-dimensional Minkowski metric by simply adding a spacelike direction. However, the general case (i.e., when $a'\ne 0$ generally but with one or more $a'=0$) would run into difficulty. For example, $a(\tau)= 3 +\sin (\tau)$ contains $a'=0$. This is a nonsingular geometry, as can be seen from the components of the Riemann curvature tensor. Like the $k=-1$ case, here too, the remedy is in six dimensions. We take the six-dimensional Minkowski metric in the following coordinates, analogous to (\ref{MinkNullCone5d}):
\begin{equation}
ds^2=-dt^2+dr^2+d\rho^2+(t+r+\rho)^2(dx^2+dy^2+dz^2)\label{MinkNullCone6d}
\end{equation}
and consider the three-dimensional curve
\begin{equation}
{\cal{C}}=(b(\eta), c(\eta), f(\eta)),
\end{equation}
which leads to the equations
\begin{eqnarray}
b + c + f&=&a, \label{k0curve16d}\\
b'^2 - c'^2-f'^2&=&1.\label{k0curve26d}
\end{eqnarray}
Again, differentiating (\ref{k0curve16d}) and solving simultaneously with (\ref{k0curve26d}) for $b'$ and $c'$, one gets a nonsingular curve for any continuous and differentiable $f(\tau)$ that does not have any common critical number with $a(\tau)$:
\begin{equation}
{\cal{C}}=\left(\int \frac{1}{2}{\frac {{a'}^{2}-2\,a'f'+ 2\,{f'}^{2}+1}{a'-f'}} d\tau, \int \frac{1}{2}{\frac {{a'}^{2}-2a'f'-1}{a'-f'}} d\tau, f(\tau) \right).
\end{equation}
One has a huge freedom in defining the curve. Choosing, for example, $f(\tau)=a(\tau)-\tau$, one obtains
\begin{equation}
{\cal{C}}=\left(\int \frac{{a'}^{2}-2a' +3}{2} d\tau,\, \int \frac{-{(a'-1)}^{2}}{2} d\tau \,,a(\tau)-\tau \right).
\end{equation}
Flipping the signs of $r$ and/or $\rho$, it is easy to check that one cannot solve for $b'$ and $c'$.
\\
\\
{\bf{Theorem 2.2:}} The general $k=0$ FLRW geometry is of embedding class two ($p=2$) with strictly one negative eigenvalue.
\subsection{Tensorial Analysis}
We now look for tensorial explanations of the above results. In particular, we would like to see how the most general $k=-1$ is $p=2$ while its two subclasses are separately $p=1$, and also how $k=0$ is $p=2$ when the scale factor has a local extremum.

For this, we start with the general fact that an arbitrary (codimension-two) spherically symmetric metric
\begin{equation}
ds^2=-A(\tau,r) d\tau^2+ B(\tau,r) dr^2 + C(\tau,r) (d\theta^2+\sin^2\theta d\phi^2) \label{genssmet}
\end{equation}
is at most $p=2$ (for a standard proof, see, for example, \cite{ExactSol2003}). The Schwarzschild solution, for example, is $p=2$.

It is also known that for a spacetime to be of embedding class one ($p=1$), a necessary and sufficient condition is that there exists a symmetric tensor $\Omega_{\mu\nu}$ satisfying (see, for example, \cite{EisenhartRiem}):
\begin{eqnarray}
R_{\mu\nu\rho\sigma}&=&\pm (\Omega_{\mu\rho} \Omega_{\nu\sigma} - \Omega_{\mu\sigma} \Omega_{\nu\rho})\label{p1condition0}\\
\Omega_{\mu\nu;\sigma}&=&\Omega_{\mu\sigma;\nu}\label{p1condition}
\end{eqnarray}
where the sign in the first equation is suitably chosen. The first equation implies the second if the determinant of $\Omega_{\mu\nu}$ is nonzero.\footnote{One immediately sees that constant curvature (i.e.\ maximally symmetric) spaces are $p=1$ with $\Omega_{\mu\nu}=g_{\mu\nu}$ satisfying (\ref{p1condition0})--(\ref{p1condition}).}

The conditions under which a spherically symmetric metric (\ref{genssmet}) becomes $p=1$ has also been considered (see, for example, \cite{KARMARKA1947}). This results in the following necessary condition:
\begin{equation}
R_{1414}R_{2323}= R_{1212}R_{3434}-R_{2142}R_{3143}.\label{p1condss}
\end{equation}
Now the general FLRW metric (\ref{CosmoCustomary}) is a special case of the spherically symmetric metric (\ref{genssmet}) with
\begin{eqnarray*}
A(\tau,r) &=& 1 \nonumber\\
B(\tau,r) &=& \frac{ a(\tau)^2}{1-kr^2} \label{sphericaltoFLRW}\\
C(\tau,r)&=& a(\tau)^2 r^2, \nonumber
\end{eqnarray*}
and its Riemann curvature tensor has six (four algebraically independent) nonzero components, Eqs (\ref{riem1212})--(\ref{riem3434}) below.
Note that these easily satisfy (\ref{p1condss}) for all $k$ and $a(\tau)$, which may have reinforced the narrative that all FLRW models are of class one.

However, (\ref{p1condss}) is only a necessary condition. So, we turn to (\ref{p1condition0})--(\ref{p1condition}) to derive $\Omega_{\mu\nu}$. It is quite straightforward to so for the FLRW geometries.\footnote{For the general spherically symmetric case, see \cite{PandeyKansal1967}.} The vanishing 14 components of the Riemann tensor imply that the off-diagonal components of $\Omega_{\mu\nu}$ are all zero. For the nonvanishing parts, one obtains
\begin{eqnarray}
R_{1212}&=& e\, \Omega_{11}\,\Omega_{22}={a a''}/{(1-kr^2)},\label{riem1212}\\
R_{1313}&=& e\, \Omega_{11}\,\Omega_{33}=aa'' r^2, \\
R_{1414} &=& e\, \Omega_{11}\,\Omega_{44}=aa'' r^2\, {\sin^2{\theta}},\\
R_{2323} &=& e\, \Omega_{22}\,\Omega_{33}= -{r^2a^2(a'^2+k)}/{(1-kr^2)},\\
R_{2424} &=& e\, \Omega_{22}\,\Omega_{44}= -{r^2a^2(a'^2+k)\,{\sin^2{\theta}}}/{(1-kr^2)},\\
R_{3434} &=& e\, \Omega_{33}\,\Omega_{44}= - a^2r^4(a'^2+k)\, {\sin^2{\theta}},\label{riem3434}
\end{eqnarray}
where $e=\pm$ of (\ref{p1condition0}). Note, again, that only four of the above are algebraically independent and they lead to the following solutions:
\begin{eqnarray}
e \Omega^2_{11}&=& -{a''}^2/(a'^2+k),\label{omega11}\\
e \Omega^2_{22}&=& -a^2 (a'^2+k)/{(1-kr^2)^2},\label{omega22}\\
e \Omega^2_{33}&=& -a^2r^4(a'^2+k),\label{omega33}\\
e \Omega^2_{44}&=& -a^2r^4(a'^2+k)\, {\sin^2{\theta}}.\label{omega44}
\end{eqnarray}
One needs to choose $e$ to make $\Omega^2_{\mu\nu}$'s positive simultaneously.

\paragraph{$k=-1$}: One needs $e=-1$ if $a'^2>1$ and $e=1$ if $a'^2<1$, which again shows the clear demarcation between the two subclasses. It is clear that, for $k=-1$, Eq. (\ref{omega11}) breaks down at $a'^2=1$ and the tensor necessary and sufficient for (\ref{CosmoCustomary}) to be $p=1$ does not exist on those points of the manifold. Thus, the most general $k=-1$ geometries are necessarily $p=2$. Note that this obstruction does not appear at the level of the Riemann tensor, which is well defined at $a'^2=1$, showing, again, that nothing is wrong with such $k=-1$ geometries.

\paragraph{$k=0$}: One needs $e=-1$ and Eq. (\ref{omega11}) breaks down at $a'^2=0$, showing the nonexistence of the tensor.

\paragraph{$k=1$}: One needs $e=-1$, and $\Omega_{\mu\nu}$ is well defined, thus they are uniformly $p=1$. (Note that for $k=1$, $r=0$ is a coordinate singularity of (\ref{CosmoCustomary}), which disappears in the other coordinates we used.)

Thus, we obtain absolutely identical results from the tensorial analysis above as we did from our curve construction. The curve construction, in addition, tells us about the signatures of the embedding spaces, which tensors are unable to do.
\subsection{Ricci flatness and Kasner's theorem}
We end this section by addressing Kasner's classic theorem, which may have come to the mind of the reader \cite{Kasner1921a}. It forbids embedding of any Ricci-flat geometry in a one-dimensional higher flat space. It follows from the (independent) components of the Ricci tensor of (\ref{CosmoCustomary}) -- $R_{11}=3a''/a, R_{22}=-({a a''+ 2 a^2 +k})/({1-kr^2})$ -- that the Ricci-flat Lorentzian FLRW geometries are necessarily flat: $a(\tau)=1$ for $k=0$ and $a(\tau)=\tau$ for $k=-1$. Thus, there is no contradiction that the large part of the FLRW is of class one. In addition, the reason for the existence of class-two geometries is not due to Ricci flatness either.
\section{Embedding FLRW in anti-de Sitter space}
We will now apply the above ideas to construct curves for $d$-dimensional FLRW geometries in a $(d+1)$-dimensional anti-de Sitter spacetime. Note that we are interested in just the geometric embedding in AdS unlike in various brane-world scenarios, where additional considerations, like brane tensions, come into play (see, for example, \cite{Nojiri, Padilla:2002tg}).

Thanks to its $SO(d-1,2)$ symmetry, the AdS$_{d+1}$ metric can be written in three different choices of static coordinates (see, for example, \cite{Gibbons:2011sg}):
\begin{equation}
ds^2=-(k+\frac{r^2}{l^2})dt^2+\frac{dr^2}{k+\frac{r^2}{l^2}}+r^2 d\Omega^2_{d-1,k}, \label{adsstaticind}
\end{equation}
where $k=-1,0,1$ and $l^2=-d(d-1)/2\Lambda$ and the other notation is as in (\ref{FLRWCos})--(\ref{FLRWCos1}). The curve
\begin{equation}
{\cal{C}}=(b(\tau), a(\tau))
\end{equation}
in the $t$--$r$ subspace (which is not flat) corresponds to the following $d$-dimensional metric:
\begin{equation}
ds^2=- d\tau^2 + a(\tau)^2 d\Omega^2_{d-1,k}
\end{equation}
provided
\begin{equation}
-(k+a^2)b'^2+\frac{a'^2}{k+a^2}=-1.\label{condads}
\end{equation}
Note that this curve is not unit speed. Rearranging (\ref{condads}), one obtains
\begin{equation}
\displaystyle b'^2=\frac{k+\frac{a^2}{l^2}+a'^2}{(k+\frac{a^2}{l^2})^2},
\end{equation}
which gives
\begin{equation}
{\cal{C}}=\left(\int \sqrt{\frac{k+\frac{a^2}{l^2}+a'^2}{(k+\frac{a^2}{l^2})^2}} d\tau, a(\tau) \right).\label{curveads}
\end{equation}
For $k=1$, both the numerator and the denominator in the integrand are strictly positive; thus, the curve exists for possible functional forms of $a(\tau)$. For $k=0$, the curve exists for all $a(\tau)\ne 0$ ($a(\tau)= 0$ reflects a breakdown of the static coordinates). On the other hand, for $k=-1$, one requires $a^2/l^2+a'^2\ge 1$ and $a(\tau)\ne l$. The condition $a(\tau)\ne l$ leaves one to consider either $0 \le a(\tau)<l$ or $a(\tau)>l$ for all $\tau$. This is, interestingly, analogous to the pseudo-Euclidean case with the difference that the demarcation is now in terms of the scale factor itself rather than its derivative. However, for $a(\tau)>l$, the other condition, $a^2/l^2+a'^2 \ge 1$, is automatically satisfied. For $a(\tau)<l$, one still has $a^2/l^2+a'^2\ge 1$ as an extra condition. For the standard cosmological models of general relativity, the latter can be eliminated, again by using (\ref{EFF}), leaving $a(\tau)<l$ as the only condition. Thus, depending on the geometry one wants to embed, $l$ can be adjusted by choosing the cosmological constant $\Lambda$. For example, any wormhole solution with an arbitrary throat radius can be embedded in a suitable AdS$_{d+1}$.
\\
\\
{\bf{Proposition 3.1:}} The following $d$-dimensional FLRW geometries are embeddable in AdS$_{d+1}$: all $k=1$, $k=0$ with $a(\tau) > 0$, $k=-1$ with $a(\tau) > l$, and $k=-1$ with $a(\tau)<l, \,a^2+a'^2\ge l$ where $l=\sqrt{-d(d-1)/2\Lambda}$.
\\
\\
Again, the last condition can be eliminated by using the Einstein equations. We do not address the questions of whether the converse is true or not and whether one can do better by moving into higher dimensional anti-de Sitter spacetimes, as we did for the pseudo-Euclidean case. These would require a more careful analysis and would take us far from the immediate use of the curve picture that we wanted to demonstrate here. As illustrations, we show below that the $d$-dimensional constant curvature spaces fall within the purview of Proposition 3.1 in more than one set of coordinates and are embeddable in AdS$_{d+1}$. We will take $l=1$ for simplicity.
\subsubsection*{Minkowski in AdS$_{d+1}$}
Minkowski can be seen as both $k=-1$ and $k=0$ FLRW. For $k=-1$, its metric is (the $d$-dimensional version of (\ref{MinkLoCone5d}))
\begin{equation}
ds^2=-d\tau^2+\tau^2 \Omega^2_{d-1,-1} \label{minkowskiddimknegative}
\end{equation}
and it corresponds to the following curve in the $t$--$r$ subspace of (\ref{adsstaticind}):
\begin{equation}
{\cal{C}}=\left(t(\tau), r(\tau) \right)=\left(\ln\sqrt{\tau^2-1}, \tau \right).\label{curveadsofmink}
\end{equation}
This is the same as the $t=\ln\sqrt{r^2-1}$ hypersurface in (\ref{adsstaticind}). Given the simplicity, it can be obtained directly as a curve $t=F(r)$ and by requiring it to give (\ref{minkowskiddimknegative}), as in the very first example in this paper with de Sitter. Note that, as expected, this hypersurface is undefined at $r=1$. Minkowski as $k=0$ is trivial with $a(\tau)=1$, giving the curve ${\cal{C}}=\left(\tau, 1\right)$.
\subsubsection*{dS$_d$ in AdS$_{d+1}$}
As we have seen, the de Sitter metric can be written in coordinates that make it $k=0$ and $k=1$. For $k=1$, we have $a(\tau)=\cosh(\tau)$, which gives the curve
\begin{equation}
{\cal{C}}=\left(t(\tau), r(\tau) \right)=\left(\frac{\sqrt{2}}{\cosh(\tau)}, \cosh(\tau) \right),\label{curveinadsofdeSitask1}
\end{equation}
and for $k=0$, we have $a(\tau)= e^{\tau}$, giving
\begin{equation}
{\cal{C}}=\left(t(\tau), r(\tau) \right)=\left(\frac{\sqrt{2}}{e^{\tau}}, e^{\tau} \right).\label{curveinadsofdeSitask0}
\end{equation}
In either case, it is the $t=\frac{\sqrt{2}}{r}$ hypersurface, as can be obtained directly.
\subsubsection*{AdS$_{d}$ in AdS$_{d+1}$}
As before, the AdS metric with $k=-1$ can be written as
\begin{equation}
ds^2=-d\tau^2+\cos^2(\tau) d\Omega^2_{d-1,-1}, \label{adsknegagain}
\end{equation}
which is a model with $a(\tau)=\cos(\tau)$. This would correspond to the curve:
\begin{equation}
{\cal{C}}=\left(t(\tau), r(\tau) \right)=\left(\mathrm{const.}, \cos(\tau) \right).\label{curveinadsofdeSitask0}
\end{equation}
This is the $t=\mathrm{const.}$ hypersurface (for $k=-1$) of (\ref{adsstaticind}),
\begin{equation}
ds^2=-\frac{dr^2}{1-r^2}+r^2 d\Omega^2_{d-1,-1},
\end{equation}
which transforms into (\ref{adsknegagain}) with $r=\cos{\tau}$.
\section{Conclusion}
The embedding of FLRW geometries has been studied independently by many researchers for over eight decades now, with every single study concluding that they are of embedding class one (even when more than one negative eigenvalue is required and the Einstein equations are   not satisfied i.e.,  the $a'<1$ sector of $k=-1$).
Given the maximal symmetry of the constant-time hypersurfaces, it is, perhaps, no surprise that only one extra dimension is needed, which also, perhaps, explains the remarkable agreement and, often a \emph{d\'{e}j\`{a}-vu}-like resemblance, between different works.
Recently,  these works have been summarized, compared, and expanded in \cite{GulamovSmolyakov2011}. It is, therefore, important to isolate the new results we added to FLRW embedding in this paper.

First, we showed that FLRW embedding in pseudo-Euclidean spaces is equivalent to unit-speed curves. With hindsight, for $k=1, -1$, this picture was already hidden within the known embedding formulae and, in particular, in the equations obtained in \cite{GulamovSmolyakov2011}, which used the same coordinates as we did for $k=1, -1$. What was left for us was to recognize the unit-speed parametrized curves and that they describe the embedding of the whole FLRW geometry they arise from. For the $k=0$ class, which Robertson described as ``surprisingly complicated compared with" the $k=1,-1$ classes, the situation was different. However, using a relatively quaint form of the Minkowski metric, (\ref{MinkNullCone5d}), we obtained the same picture, and a new set of embedding formulae, for $k=0$.

Recall that the embedding of the $d$-dimensional de Sitter spacetime as a hyperboloid in $(d+1)$-dimensional Minkowski space, suppressing dimensions, is often presented as a two-dimensional hyperboloid in a three-dimensional Minkowski space. FLRW cosmologies, with their isotropic parts suppressed, can similarly be seen as embedded surfaces in the three-dimensional Minkowski space (see, for example, \cite{LachiezeRey:2000my}). What we have found here is that by using appropriate matching coordinates, the picture can further be reduced to that of a unit-speed curve. While it is simpler to see this for de Sitter (cf.\ our first example), for FLRW geometries, this emerges naturally only if one uses a parametrized description of the curves.

In particular, the example of $k=0$ taught us that, starting with matching coordinates, the existence of the unit-speed parametrized curve can be turned into a probe for determining the embedding. This lesson was immediately put to use, as we noticed two missing possibilities within the $k=0$ and $k=-1$ classes. Constructing curves in appropriate coordinates, we proved that the most general $k=-1$ and $k=0$ geometries in four dimensions -- which allow the missing possibilities -- are of embedding class two, i.e., minimally embeddable in six dimensions with appropriate signatures, which stand in sharp contrast to the unanimous consensus on five dimensions. This completes the mathematical picture of FLRW embedding, and may be physically useful in other theories of gravity. We then looked at defining appropriate tensors, which produced conclusions identical to those obtained from the curve picture, and demonstrated the latter's rigor (as well as the particular advantage of being able to determine the signature of the embedding).

Note that the different coordinates that we used for the Minkowski metric do not cover the whole Minkowski space as the canonical coordinates do. However, this poses no problem since each FLRW model is embedded within whatever part of the Minkowski space the respective coordinates cover. These coordinates can be transformed back to the canonical ones by standard prescriptions; however, there is no need to do so since the vanishing of the Riemann tensor guarantees that any line element can be transformed into the canonical Cartesian coordinates with the same numbers of positive and negative eigenvalues (see, for example, \cite{EisenhartRiem}).

As the next natural thing, using curve construction, we considered embedding in AdS space. We found that, quite remarkably, most of the FLRW models can be embedded in AdS space in one higher dimension, with almost comparable success as in the Minkowski space. In fact, in the limit of $l\rightarrow \infty$, one recovers the pseudo-Euclidean results for $k=1,-1$, but not for $k=0$. This shows again the special nature of the null-cone metric (\ref{MinkNullCone5d}). It would be interesting if there were a single Ricci-flat space, other than the flat space, that would embed all or most of the FLRW models, like the pseudo-Euclidean and the AdS spaces do.
\section*{Acknowledgements}
We thank Malcolm MacCallum for the discussion on nomenclature mentioned in footnote 3.


\vskip1cm
\begin{thebibliography}{99}

\bibitem{HE}
S.~W.~Hawking and G.~F.~R.~Ellis (1973).
\emph{The Large Scale Structure of Space-Time}
(Cambridge University Press, UK).

\bibitem{Besse}
A. Besse (1986). \emph{Einstein Manifolds} (Springer, Berlin).

\bibitem{EisenhartRiem}
L. Eisenhart (1925).
\emph{Riemannian Geometry} (Princeton University Press, New Jersey)

\bibitem{Friedman1965}  A. Friedman, ``Isometric Embeddings of Riemannian Manifolds into Euclidean Spaces," Rev.
Mod. Phys. \textbf{37} (1965) 201.

\bibitem{Pavsic:2000qy}
  M.~Pavsic and V.~Tapia,
  ``Resource  Letter on Geometrical Results for Embeddings and Branes,''
  gr-qc/0010045.

\bibitem{ExactSol2003}
 H.~Stephani, D.~Kramer, M.~A.~H.~MacCallum, C.~Hoenselaers and E.~Herlt (2003).
   \emph{Exact Solutions of Einstein's Field Equations} (Cambridge University Press, UK).

\bibitem{Robertson:1928}  H.~P.~Robertson,
  ``On Relativistic Cosmology,'' Philos. Mag. {\bf{5}}  (1928) 835--848

\bibitem{Robertson:1933zz}
  H.~P.~Robertson,
  ``Relativistic Cosmology,''
  Rev.\ Mod.\ Phys.\  {\bf 5} (1933) 62.
  
\bibitem{Rosen1965} J. Rosen, ``Embedding of Various Relativistic Riemannian Spaces in Pseudo-Euclidean Spaces" Rev. Mod. Phys. {\bf{37}}  (1965) 204--214

\bibitem{GulamovSmolyakov2011}
I. E. Gulamov and M. N. Smolyakov,
``Submanifolds in Five-dimensional Pseudo-Euclidean
Spaces and Four-dimensional FRW Universes,'' Gen. Rel. Grav. {\bf{44}} (2012), 703--710

\bibitem{Nielsen} B. Nielsen ``Minimal Immersions, Einstein's Equations and Mach's Principle,"
 J. Geom. Phys. \textbf{4} (1987), no. 1, 1--20
 
\bibitem{Paston:2012pb}
  S.~A.~Paston and A.~A.~Sheykin,
  ``Embeddings for Schwarzschild Metric: Classification and New Results,"
  Class.\ Quant.\ Grav.\ {\bf 29}, 095022 (2012)


\bibitem{Andrianopoli:1999kx}
 L.~Andrianopoli, M.~Derix, G.~W.~Gibbons, C.~Herdeiro, A.~Santambrogio and A.~Van Proeyen,
  ``Isometric Embedding of BPS Branes in Flat Spaces with Two Times,''
  Class.\ Quant.\ Grav.\  {\bf 17} (2000) 1875
  [arXiv:hep-th/9912049].


\bibitem{DoCarmo}
M. P. Do Carmo (1976). \emph{Differential Geometry of Curves and Surfaces} (Eaglewood Cliffs, NJ: Prentice-Hall)

\bibitem{ONeill1}
B. O'Neill (2006).
\emph{Elementary Differential Geometry} (Academic Press, New York).

\bibitem{ONeill}
B. O'Neill (1983).
\emph{Semi-Riemannian Geometry with Applications to Relativity} (Academic Press, New York).

\bibitem{WaldBook}
R. M. Wald (1984) \emph{General Relativity} (University of Chicago Press, Chicago).

\bibitem{LachiezeRey:2000my}
  M.~Lachi\`{e}ze-Rey,
  ``The Friedmann-Lema\^{\i}tre Models in Perspective: Embeddings of the
  Friedmann-Lema\^{\i}tre models in Flat 5-Dimensional Space,''
  Astron.\ Astrophys.\  {\bf 364} (2000) 894
  [arXiv:astro-ph/0010163].

\bibitem{Robinson:1987mr}
  I.~Robinson,
  ``On Plane Waves and Nullicles," in
  \emph{From SU(3) to Gravity: Festschrift in
Honor of Yuval Ne'eman (Eds E Gotsman, G Tauber)} (1985) (Cambridge University Press, UK), 409--422.

\bibitem{KARMARKA1947}
K. R. Karmarkar
``Gravitational Metrics of Spherical Symmetry and Class One," \emph{Proc. Indian Acad. Sci. A} \textbf{27}, 56.

\bibitem{PandeyKansal1967} S. N. Pandey and I. D. Kansal, ``Spherically Symmetric Space-time of Class One and Electromagnetism,"
\emph{Proc. Cambridge Philos. Soc.} \textbf{64} (1968) 757--764.

\bibitem{Kasner1921a}
E.~Kasner, ``The Impossibility of Einstein Fields Immersed in Flat Space of Five Dimensions." Amer. J. Math. \textbf{43}, no. 2 (1921) 126--29


\bibitem{Nojiri}
S.~Nojiri, S.~D.~Odintsov and S.~Ogushi,
``Friedmann-Robertson-Walker Brane Cosmological Equations from the Five-dimensional Bulk (A)dS Black Hole,''
Int.\ J.\ Mod.\ Phys.\ A {\bf 17} (2002) 4809
[arXiv:hep-th/0205187].

\bibitem{Padilla:2002tg}
  A.~Padilla,
  ``Brane  World Cosmology and Holography,''
  hep-th/0210217.


\bibitem{Gibbons:2011sg}
  G.~W.~Gibbons,
  ``Anti-de-Sitter Spacetime and Its Uses,'' Lect. Notes. Phys. \textbf{537} (2000), 102.
  arXiv:1110.1206



\end{thebibliography}
\end{document}